\documentclass[twocolumn,showpacs,pra,amsmath,amssymb,floatfix]{revtex4-1}
\input{epsf}

\usepackage{graphicx}
\usepackage{dcolumn}
\usepackage{bm}

\begin{document}
\title{Coherent control of atomic spin currents in a double well}
\author{H. T. Ng${}^{1}$ and Shih-I Chu${}^{1,2}$}
\affiliation{${}^{1}$Center for Quantum Science and Engineering, Department of Physics, National Taiwan University, Taipei 10617, Taiwan}
\affiliation{${}^{2}$Department of Chemistry, University of Kansas, Lawrence, Kansas 66045, USA}
\date{\today}

\begin{abstract}
We propose a method 
for controlling the atomic currents of a 
two-component Bose-Einstein condensate in 
a double well by applying an external field 
to the atoms in one of the potential wells. 
We study the ground-state properties of the 
system and show that the directions of spin currents 
and net-particle tunneling can be manipulated 
by adiabatically varying the coupling strength
between the atoms and the field. 
This system can be used for studying spin and 
tunneling phenomena across a wide range of 
interaction parameters. In addition, 
spin-squeezed states can be 
generated. It is useful for 
quantum information processing and quantum metrology.
\end{abstract}


\maketitle

\section{Introduction}
Spin and tunneling phenomena are of fundamental 
interests in understanding quantum behaviors of 
particles. They are also important in the applications 
using solid-state devices such as sensors and data 
storage \cite{Fert}. In addition, manipulation of 
the quantum state of a single spin 
is essential in implementing quantum information 
processing \cite{Nielsen}.

Recently, the tunneling dynamics of 
ultracold atoms has been observed in a double well \cite{Albiez}
and optical lattices \cite{Folling}, 
where the experimental parameters can 
be widely tuned. Moreover, high-fidelity 
single-spin detection of an atom has been 
realized in an optical lattice \cite{Bakr,Weitenberg} 
and atom-chip \cite{Volz}, respectively.
Such sophisticated techniques of manipulating 
ultracold atoms open up the possibilities for 
the study of intriguing quantum phenomena never 
possible before.

In this paper, we propose a method to control 
the tunneling dynamics of a two-component 
Bose-Einstein condensate (BEC) in a double well \cite{Ng}
by applying an external field to the atoms in 
one of the potential wells. In fact, the methods for 
controlling of tunneling in a double-well BEC have 
been suggested by driving the double-well potentials \cite{Holthaus}
and by applying a symmetry-breaking field \cite{MMolina},
respectively.
Besides, a number of methods for manipulating the atomic motions in  
an an optical lattice have been proposed such as 
using external fields for vibrational 
transitions between adjacent sites \cite{Forster}, 
tilting the lattice potential \cite{DFrazer}
and periodic modulation of the lattice parameters \cite{Creffield}.

Here we show that the spin and tunneling dynamics 
of atoms can be manipulated by adiabatically changing 
the coupling strength of the field. This approach 
can be used for studying the spin and tunneling related 
phenomena in a controllable manner. For example, 
the directions of ``spin currents'' such as {\it parallel-} 
and {\it counter-}flows can be controlled by 
appropriately adjusting the interaction parameters. 
Here the ``spin currents'' refer to the two atomic 
currents of a condensate of ${}^{87}$Rb atoms with the 
two different hyperfine levels \cite{Harber}. 
Apart from changing the internal states of atoms, 
the external field gives rise to 
{\it net-particle tunneling}. It is different to 
the situation of co-tunneling of the two component 
condensates in a double well \cite{Ng,Kuklov}
in which the number 
difference of atoms between the wells is equal 
during the tunneling process.

In addition, the tunnel behaviors are 
totally different in the limits of weak and 
strong atomic interactions. In the regime 
of weak atomic interactions, the atoms smoothly 
tunnel through the other well. On the contrary, 
the discrete steps of tunneling are shown in the 
limit of strong atomic interactions. The coherent control 
of single-atom tunneling can thus be achieved. 
This method can be utilized for atomic transport.

Furthermore, we investigate the production of 
spin squeezing \cite{Kitagawa,Wineland} in this system. 
The occurrence of spin-squeezing can indicate 
multi-particle entanglement \cite{Sorensen}. We show that 
spin-squeezed states can be dynamically 
generated by slowly changing the coupling
strength of the field. This can be
used for preparing entangled states and
quantum metrology \cite{Esteve}. 

This paper is organized as follows:
In Sec.~II, we introduce the system
of a two-component BEC in a double
well, where the atoms in the left 
potential well are coupled to the
laser field. In Sec.~III,
we study the ground-state properties
of the system. In Sec.~IV, we propose
a method to adiabatically transfer 
the atoms to the other well. In Sec.~V,
we investigate the generation of 
multi-particle entanglement using this
method. We provide a summary in Sec.~VI.
In the Appendix, we derive an effective
Hamiltonian describing the coupling
between the two internal states with
the lasers.

\section{System}
\begin{figure}[ht]
\centering
\includegraphics[height=3cm]{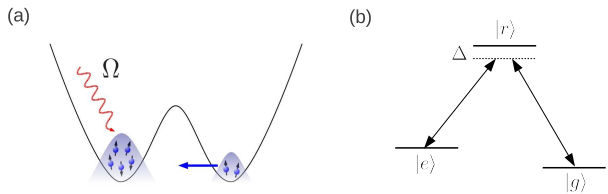}
\caption{ \label{DW_fig1} (Color online) (a) Schematic
of a two-component Bose-Einstein condensate in a double well.
An external field is applied to the atoms in the left
potential well.  (b) Energy levels for the atoms.  
The two internal states $|e\rangle$ and $|g\rangle$
are coupled via the upper state $|r\rangle$.
}
\end{figure}
We consider a condensate of ${}^{87}$Rb atoms 
with two hyperfine levels 
$|e\rangle=|F=2,m_F=1\rangle$ and $|g\rangle=|F=1,m_F=-1\rangle$ 
of the $5S_{1/2}$ ground state \cite{Harber}
confined in a 
symmetric double-well potential \cite{Ng}. An external
field is applied to the atoms in one of the potential wells.
The schematic of the system is shown in Fig.~\ref{DW_fig1}(a).
This system can be described by the total Hamiltonian 
$H_{\rm tot}=H_{0}+H_{1}$, where $H_{0}$ 
and $H_{1}$
are the Hamiltonians describing the external and internal
degrees of atoms.

We adopt the two-mode approximation to describe 
the atoms in deep potential wells \cite{Ng}.  Since the scattering
lengths of the different hyperfine states of ${}^{87}$Rb
are very similar \cite{Matthews}, we assume that the intra- and inter-component
interactions are nearly the same.
The Hamiltonian $H_{0}$ can be written as ($\hbar=1$)
\begin{eqnarray}
\label{Hamext}
 H_{0}&=&-J(e^{\dag}_{L}e_{R}+e^\dag_{R}e_{L}+g^{\dag}_{L}g_{R}+g^\dag_{R}g_{L})\nonumber\\
&&+U[({n}_{eL}+n_{gL})^2+(n_{eR}+n_{gR})^2],
\end{eqnarray}
where $\alpha_{L}(\alpha_R)$ and 
$n_{\alpha_{L}}(n_{\alpha_R})$ are annihilation operator of an atom
and number operator in the left(right) potential well,
for $\alpha=e,g$. The total number $N$ of atoms 
is conserved in this system. To ensure the validity of
two-mode approximation, we assume that the trapping energy
is much larger than the atomic interaction energy \cite{Milburn,Holthaus}, i.e.,
$\omega_0\gg{UN}$, where $\omega_0$ is the effective trapping
frequency of the potential well.

We can make a rough estimation of the experimental
parameters within the two-mode approximation.
For ${}^{87}$Rb, the scattering length $a$ is about
5 nm. We take the frequencies $\omega_{y}$ and $\omega_z$ of 
the transervse trapping potentials to be about $2\pi{\times}1$ kHz.
Indeed, the barrier height and the separation between the
two potential minima can be varied \cite{Maussang} by appropriately
splitting the potential. We consider that the barrier height
$V_b$ can be tuned \cite{Maussang} from $2\pi\times50$ to 
$2\pi\times250$ Hz and the separation
$2x_0$ between two wells are 2 and 4 $\mu$m. The effective
frequency $\omega_0=\sqrt{8V_b/(mx^2_0)}$ \cite{Milburn}
is related to the barrier height $V_b$ and the separation $2x_0$. 
This gives the effective frequency $\omega_0$ 
ranging from $2\pi\times{220}$ to $2\pi\times{350}$ Hz,
and the interaction strength $U$ ranges between 
$2\pi\times2.7$ and $2\pi{\times}3.4$ Hz.
From these estimations, the number $N$ of atoms must be
less than 100 to maintain the validity of two-mode
approximation \cite{Milburn,Holthaus}. We also estimate 
the ratio $U/J$ of the atomic strength and tunneling 
strength which ranges from 0.05 to 300.

Without loss of generality, we consider an external field 
to be applied to the atoms in the left potential well.
The two internal states can be coupled via the upper
transition of the ${\rm D}_2$ line of ${}^{87}$Rb \cite{Steck}
by using the two laser beams with different circular polarizations
as shown in Fig.~\ref{DW_fig1}(b).  
This upper state $|r\rangle$ can be adiabatically eliminated
due to the large detuning. In the Appendix A, we derive an effective
Hamiltonian for describing the interaction between the two 
internal states and the lasers. 
In the interaction picture, the 
Hamiltonian $H_{1}$ is given by \cite{Barnett}
\begin{eqnarray}
\label{Hamint}
 H_{1}&=&{\Delta}(n_{eL}+n_{eR})
+{\hbar}\Omega(e^\dag_{L}g_{L}+{\rm H.c.}),
\end{eqnarray}
where $\Delta$ is the detuning between the atomic transition
($|r\rangle$ and $|e\rangle$) and the laser field, 
and $\Omega$ is the effective coupling 
strength between the atoms and external field.   
A tightly focused laser can be applied to
the atoms in one of the potential well. In fact,
a tightly focused laser has been used for 
addressing a single atom in an optical lattice
\cite{Weitenberg}, where the full-width
at half-maximum (FWHM) of the diameter of 
laser beam is within 1 $\mu$m \cite{Weitenberg}. 
The separation between the two potential wells
is about several $\mu$m in the experiment \cite{Albiez}.
Therefore, the effects of the external lasers on
the atoms in the other well are small.

\section{Ground-state properties of the coupled atom-laser system}
\begin{figure}[ht]
\centering
\includegraphics[height=6.5cm]{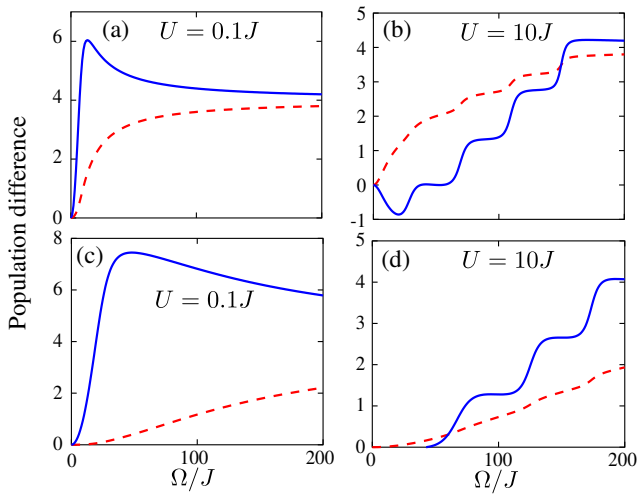}
\caption{ \label{GSpdiff_spcom_D_20J_D_200J_U_01J_U_10J} (Color online) 
Population differences are plotted as a function 
of $\Omega/J$ for the ground state of the system and $N=8$. The parameters
are shown: (a) and (b) $\Delta=20J$;
(c) and (d) $\Delta=200J$.
The ground state $|g\rangle$ and excited state $|e\rangle$ 
of atoms are denoted by blue-solid and
red-dotted lines, respectively.
}
\end{figure}

Now we study how the ground state properties
of the BEC affected by the local
external field. 
In Fig.~\ref{GSpdiff_spcom_D_20J_D_200J_U_01J_U_10J}, we plot
the population differences 
$(\langle{n_{\alpha{L}}}\rangle-\langle{n_{\alpha_R}}\rangle)$
versus the coupling strengths $\Omega$ for the atoms in the
two different internal states $|\alpha\rangle$ and
$\alpha=e,g$. 
For the cases of even number $N$ of atoms, 
there is an equal number of atoms in
the two wells in the absence of the external 
field, i.e., $\Omega=0$. 
The external field
causes the energy bias between the two wells.
Thus, the population difference becomes larger 
when the coupling strength increases.

Moreover, the system exhibits totally different 
behaviors in the regimes of
weak and strong atomic interactions.  For weak 
atomic interactions, the population differences smoothly vary 
as a function of the coupling strength as shown in 
Figs.~\ref{GSpdiff_spcom_D_20J_D_200J_U_01J_U_10J}(a) and (c). 
Also, a larger number of atoms are 
in the state $|g\rangle$ due to the larger detuning $\Delta$.  

In Figs.~\ref{GSpdiff_spcom_D_20J_D_200J_U_01J_U_10J}(b) and (d),
we plot the population differences versus the coupling strengths
$\Omega$ in the regime of strong atomic interactions.
We can see that discrete steps of the population difference for 
atoms in the state $|g\rangle$ are shown 
when the coupling strength increases. 
However, the discrete feature is not obvious for 
the atoms in the state $|e\rangle$.  
In Fig.~\ref{GSpdiff_spcom_D_20J_D_200J_U_01J_U_10J}(b),
the atoms, in the two different internal states, 
distribute in the opposite potential wells
for small $\Omega$. When $\Omega$ becomes larger, 
both component of atoms populate in
the left potential well. This result shows that 
the population difference of atoms in the two internal
states depends on the coupling strength
and also the detuning between the atoms and field.

To proceed, we investigate the relationship between the total
population difference of atoms ($\langle{n_{eL}}+{n_{gL}}-{n_{eR}}-n_{gR}\rangle$)
and the coupling strength $\Omega$. 
In Fig.~\ref{GStpdiff_D_20J_N_8_3},
we plot the total population
differences as a function of the coupling
strength $\Omega$ for different strengths $U$
of atomic interactions.
The external field 
leads to the population imbalance between
two wells in both regimes of 
weak and strong atomic interactions.  
For weak atomic interactions,
the population difference smoothly increases
with $\Omega$.  When the atomic interactions 
become strong, the discrete 
steps of population differences 
are shown in Fig.~\ref{GStpdiff_D_20J_N_8_3}. 
The sharper discrete steps are shown for larger $U$.  
In this case, a single atom 
is only allowed to tunnel through the other well for 
the specific coupling strengths.

\subsection{Tunneling condition in the limit of strong atomic interactions}
We now discuss the tunneling condition in the limit of 
strong atomic interactions.  
Since the tunnel couplings are negligible in this regime,
the numbers of atoms in the two wells are conserved.
We assume that there are $N/2+n$ and $N/2-n$ atoms
in the left and right wells.  
For convenience, we define the angular momentum operators:
$S_{jx}=(g_je^\dag_j+e_jg^\dag_j)/2$,
$S_{jy}=(g_je^\dag_j-e_jg^\dag_j)/2i$ and
$S_{jz}=(e^\dag_je_j-g^\dag_jg_j)/2$, and $j=L,R$.
\begin{figure}[ht]
\centering
\includegraphics[height=5cm]{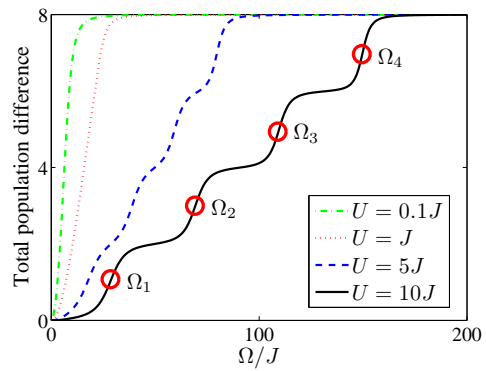}
\caption{ \label{GStpdiff_D_20J_N_8_3} (Color online) 
Plot of the total population difference versus $\Omega/J$
for the ground state with the different strengths $U$ of 
atomic interactions. The parameters are used: $N=8$ and $\Delta=20J$.
The values of $\Omega_n$ in Eq.~(\ref{genergy}) are
marked with the red empty circles.
}
\end{figure}
To diagonalize the Hamiltonian $H_{\rm tot}$, we apply the transformation
as 
\begin{eqnarray}
S_{Lx}&=&\cos\theta{S_{Lx}'}-\sin\theta{S_{Lz}'},\\
S_{Lz}&=&\cos\theta{S_{Lz}'}+\sin\theta{S_{Lx}'}.
\end{eqnarray}
By setting the term, $\Delta\sin\theta+2\Omega\cos\theta$, to 
zero, the total Hamiltonian can be diagonalized as
\begin{eqnarray}
H_{\rm tot}'&=&{\Delta}S_{Rz}+\hbar\sqrt{\Delta^2+4\Omega^2}S_{Lz}'
+U(N^2/2+2n^2)\nonumber\\
&&+\hbar\Delta{N}/2,
\end{eqnarray}
and its ground-state energy $E^{G}_n$ is given by
\begin{eqnarray}
\label{genergy}
E^{\rm G}_n&=&-\Delta(N/2-n)/2-\sqrt{\Delta^2+4\Omega^2}(N/2+n)/2\nonumber\\
&&+2{U}n^2+UN^2/2+\Delta{N}/2.
\end{eqnarray}
The tunneling of atoms occurs when the energies
$E^{\rm G}_n$ and $E^{\rm G}_{n-1}$ are equal to each other.
In this situation, a single atom tunnels through 
the other well. It is analogous to the resonant tunneling
in quantum dots due to the Coulomb blockade \cite{Beenakker}.
By setting $E^{\rm G}_n=E^{\rm G}_{n-1}$ in Eq.~(\ref{genergy}), 
the tunneling condition for the coupling strength $\Omega_n$ 
can be obtained as
\begin{eqnarray}
\label{tunnel_cond}
\Omega_n&=&\frac{1}{2}\{[4U(2n-1)+\Delta]^2-\Delta^2\}^{\frac{1}{2}}.
\end{eqnarray}
In Fig.~\ref{GStpdiff_D_20J_N_8_3}, the values of $\Omega_n$
are marked with the red empty circles. This shows that
the tunneling condition in Eq.~(\ref{tunnel_cond}) 
agrees with the exact numerical solution.

\section{Adiabatic transport}
We have studied the ground-state properties
of the coupled atom-laser system in the previous
section. We have shown that the directions of
atomic currents and the population difference
depend on the coupling strength $\Omega$.
Now we study the adiabatic transport of atoms by
slowly increasing the coupling strength 
$\Omega$. According to the adiabatic theorem \cite{Messiah}, 
the system can evolve as its instantaneous 
ground state if the changing rate of
the parameter $\Omega$ is sufficiently slow. 
Therefore, this method can be used for 
adiabatically transferring the atoms to 
the other potential well.
\begin{figure}[ht]
\centering
\includegraphics[height=6.0cm]{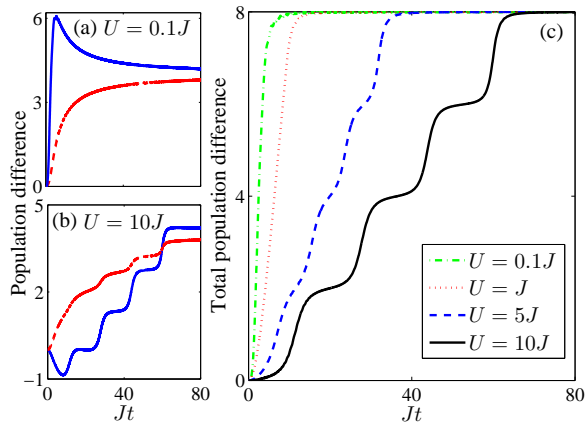}
\caption{ \label{tdep_tpdiff_D_20J_N_8_v_2d5J} (Color online) 
The population differences for the two different component condensates
are plotted against the time $Jt$ for $U=0.1J$ and $10J$ in (a)
and (b). The atoms in ground and excited states 
are denoted by blue-solid and red-dashed lines, respectively.
(c) Plot of the total population difference versus the time $Jt$ for different
strengths of atomic interactions. 
The parameters are used: $N=8$, $\Delta=20J$ and $v=2.5J$.
}
\end{figure}

We consider the coupling strength 
$\Omega(t)$ as a linear function of time $t$, i.e.,
\begin{eqnarray}
\label{Omegat}
\Omega(t)&=&v{t},
\end{eqnarray}
where $v$ is a positive number.
Here the detuning $\Delta$ is kept
as a constant during the time 
evolution.
Initially, the system is prepared in 
its ground state of the Hamiltonian $H_0$
in Eq.~(\ref{Hamext}) and setting $\Omega=0$.
The coupling strength $\Omega(t)$ in
Eq.~(\ref{Omegat}) is adiabatically increased.
In Figs.~\ref{tdep_tpdiff_D_20J_N_8_v_2d5J}(a) and (b),
the population differences are plotted
versus the time for the atoms in the 
different internal states. We can see that the 
two results 
in Figs.~\ref{GSpdiff_spcom_D_20J_D_200J_U_01J_U_10J} 
and \ref{tdep_tpdiff_D_20J_N_8_v_2d5J} 
reach a good agreement. This shows that the 
tunneling dynamics of atoms can be 
controlled by using an external field.

In the regime of weak atomic interactions,
the two different component condensates
smoothly tunnel through the other well in the 
same direction in Fig.~\ref{tdep_tpdiff_D_20J_N_8_v_2d5J}(a).  
In the opposite interaction limit, the discrete steps
are shown in 
Fig.~\ref{tdep_tpdiff_D_20J_N_8_v_2d5J}(b). 
Note that the counter-flow is
shown in a short time in 
figure (b). The flows of atomic spin currents 
become parallel afterward.  
This shows that the direction of spin flows can 
be controlled by appropriately adjusting the parameters
$\Omega$ and $\Delta$.

Then, we study the time evolution of total population
difference of atoms.
In Fig.~\ref{tdep_tpdiff_D_20J_N_8_v_2d5J}(c), 
we examine the tunneling dynamics for a wide range of 
interaction parameters.
In both regimes, the atoms tunnel
to the left potential well when the coupling
strength slowly increases with time. 
The total population differences smoothly 
increases as a function of time for weak atomic
interactions.  The discrete
steps of the tunneling are shown when the 
strength of atomic interactions becomes strong. 
The single-atom tunneling
can thus be achieved in the limit of 
strong atomic interactions.

\begin{figure}[ht]
\centering
\includegraphics[height=6.0cm]{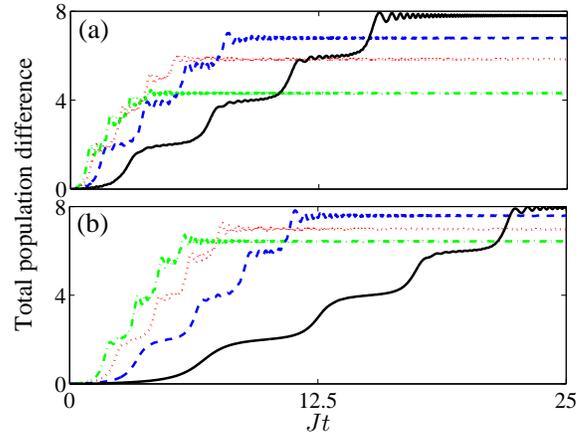}
\caption{ \label{tdep_tpdiff_D_20J_D_200J_N_8_U_10J_2} (Color online) 
The total population differences are plotted versus 
the time $Jt$ for the different detunings $\Delta=20J$ 
and $200J$ in (a) and (b), respectively.
The different rates of change are denoted: $v=10J$ (black-solid line),
$v=20J$ (blue-dashed line), $v=30J$ (red-dotted line) and
$v=40J$ (green-dashed-dotted line), respectively.
The parameters are used: $N=8$ and $U=10J$. 
}
\end{figure}
\subsection{Efficiency of the population difference}
Next, we investigate the efficiency of 
the population transfer by increasing the coupling
strength of the external field. 
In Figs.~\ref{tdep_tpdiff_D_20J_D_200J_N_8_U_10J_2}(a) 
and (b), the population differences
are plotted versus the time for the two
different detunings $\Delta=20J$ and 
$200J$.  The full population 
transfer can be achieved if the
parameters $v$ are small enough in both cases. 
When the parameters $v$ become larger,
the rates of population transfer increase.
However, the smaller population of atoms
can be transferred in both cases.
It is because they have gone beyond the adiabatic limit.
By comparing the figures \ref{tdep_tpdiff_D_20J_D_200J_N_8_U_10J_2}(a) 
and (b), we find that the larger numbers of atoms can be transported
with the same rate $v$ of change 
for the case using a larger detuning.

\subsection{Effects of the coupling between the laser
and the atoms in the neighboring well}
Since the separation between the two wells
is small \cite{Albiez,Maussang}, the lasers may also couple the 
atoms in the other potential well.
Now we examine a very small coupling between the 
laser and the atoms in the right potential well.
The Hamiltonian, describes the coupling 
between the laser and the atoms in the right potential
well, can be written as
\begin{eqnarray}
H^{\rm R}_{1}=\Omega'(e^\dag_Rg_R+g^\dag_Re_R),
\end{eqnarray}
where $\Omega'$ is the coupling strength between
the laser and atoms.
\begin{figure}[ht]
\centering
\includegraphics[height=6.0cm]{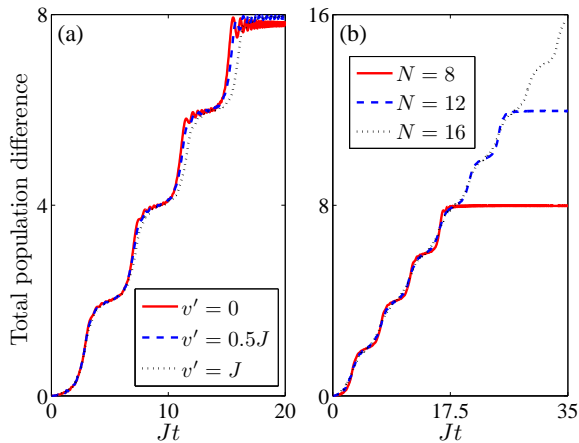}
\caption{ \label{rcoupl1} (Color online) 
Time evolution of the total population difference.
The effect of a small coupling between the laser 
and the atoms in the right potential well
is investigated. In (a), the total population 
differences are plotted versus 
the time $Jt$ for different rates of changing
$v'$ and $N=8$. In (b), the total population differences 
are plotted versus the time $Jt$ for different numbers of
atoms and $v'=J$. The parameters are used: $U=10J$, $\Delta=20J$
and $v=10J$. 
}
\end{figure}
Here we consider that the coupling strength $\Omega'$
linearly increases with time as
\begin{eqnarray}
\Omega'(t)=v't.
\end{eqnarray}
We assume that the parameter $v'$
is much smaller than $v$ in Eq.~(\ref{Omegat}).

In Fig.~\ref{rcoupl1}(a), the total population differences
are plotted versus the time for the different
rates of changing $v'$.  The atoms can be
efficiently transferred from the left to right potential 
well. The results show slightly different tunneling
behaviors for small $v'$. We then examine
the population transfer with small $v'$
for larger number of atoms.  In Fig.~\ref{rcoupl1}(b),
we plot the total population difference as
a function of time for different numbers 
of atoms.  The discrete steps
of tunneling can be clearly shown.  This shows
that this method works even if there is a very 
small coupling between the laser and 
the atoms in the neighboring well.

\section{Multi-particle entanglement}
Having discussed the tunneling dynamics of atoms,
we study the generation of spin-squeezed states
by adiabatically changing the coupling strength of the field. 
To indicate the occurrence of spin squeezing, a
parameter $\xi^2$ can be 
defined as \cite{Wineland}
\begin{eqnarray}
\xi^2&=&\frac{N(\Delta{S_{n_1}})^2}{\langle{S_{n_2}}\rangle^2+\langle{S_{n_3}}\rangle^2},
\end{eqnarray}
where $n_i$ is the $i$-th component of an angular momentum system,
and $i=1,2$ and 3. If $\xi$ is less than one, then 
the system is said to be spin-squeezed \cite{Wineland}.
In addition, the parameter $\xi$
can be used for indicating multi-particle entanglement \cite{Sorensen}.
\begin{figure}[ht]
\centering
\includegraphics[height=5cm]{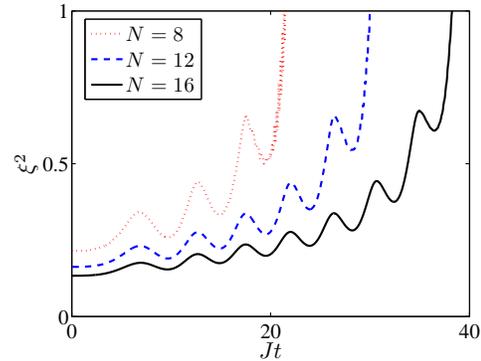}
\caption{ \label{spin_squeezing_U10J_D_200J_v_10J_2} (Color online) 
The spin-squeezing parameters $\xi^2$ are plotted versus the time $Jt$
for different number $N$ of atoms. The parameters are used: $\Delta=200J$, 
$U=10J$ and $v=10J$.  
}
\end{figure}

Let us define the angular momentum operators as:
$S_x=(e^\dag_Le_R+g^\dag_Lg_R+e^\dag_Re_L+g^\dag_Rg_L)/2$,
$S_y=(e^\dag_Le_R+g^\dag_Lg_R-e^\dag_Re_L-g^\dag_Rg_L)/2i$,
and $S_z=(e^\dag_Le_R+g^\dag_Lg_L-e^\dag_Re_R-g^\dag_Rg_R)/2$.
The angular momentum operators obey the standard commutation rule. 
We study the parameter as
\begin{eqnarray}
\xi^2&=&\frac{N(\Delta{S_{z}})^2}{\langle{S_{x}}\rangle^2}.
\end{eqnarray}
Physically speaking, the quantity $(\Delta{S}_z)^2$ is
the variance of the total population difference of atoms 
between the wells, and $\langle{S}_x\rangle$ is the sum of
the phase coherences between the two wells for the two 
component condensates.

In Fig.~\ref{spin_squeezing_U10J_D_200J_v_10J_2}, 
we plot the spin-squeezing parameter $\xi^2$
versus the time for different number of atoms. 
Initially, the parameter $\xi^2$ is below one
when $\Omega=0$. 
This means that the initial ground state is a spin-squeezed state.
As $\Omega$ in Eq.~(\ref{tunnel_cond}) slowly increases with time, 
spin squeezing can be dynamically produced. This result indicates that 
the system is spin-squeezed for a wide range of $\Omega$.
Besides, a higher degree of spin squeezing can be produced 
with a larger number of atoms $N$.

\section{Conclusions}
In summary, we have studied how the ground
state of a two-component condensate in a 
double well affected by a local external field.  
We have shown that the flows of spin currents
and particle-tunneling dynamics
can be controlled by slowly 
varying the coupling strength of the
external field and appropriately 
adjusting the detuning. 
This can be used for studying
spin and tunneling phenomena and also
the potential applications of atomic 
devices in atomtronics \cite{Pepino}.
In addition, spin-squeezed states 
can be dynamical generated. It is an
important resource for precision measurement 
\cite{Wineland,Kitagawa,Sorensen}.

\begin{acknowledgments}
This work was partially supported by 
US National Science Foundation. We would 
like to also acknowledge the partial 
support of National Science Council and 
NTU-MOE.
\end{acknowledgments}

\appendix
\section{Derivation of the effective Hamiltonian for 
coupling between two internal states and
the lasers}
The two internal states of an atom can be coupled 
via pumping to the upper state
by two lasers as shown in Fig.~\ref{DW_fig1}(b).  
The Hamiltonian, describes the interaction
between the two internal states and the lasers,
can be written as
\begin{eqnarray}
H&=&\omega_{rg}|r\rangle\langle{r}|+\omega_{eg}|e\rangle\langle{e}|
+\Omega_g[\exp{(-i\omega_g{t})}|r\rangle\langle{g}|+{\rm H.c.}]\nonumber\\
&&+\Omega_e[\exp{(-i\omega_e{t})}|r\rangle\langle{e}|+{\rm H.c.}],
\end{eqnarray}
By performing the unitary transformation as,
\begin{eqnarray}
U(t)&=&\exp\{-i[\omega_g|r\rangle\langle{r}|+(\omega_g+\omega_e)|e\rangle\langle{e}|]t\},
\end{eqnarray}
the Hamiltonian can be transformed as \cite{Barnett}
\begin{eqnarray}
H'&=&i\dot{U}^{\dag}U+U^\dag{H}U,\\
&=&\Delta_r|r\rangle\langle{r}|+\Delta_e|e\rangle\langle{e}|+(\Omega_g|r\rangle\langle{g}|+\Omega_e|r\rangle\langle{e}|+{\rm H.c.}),\nonumber\\
\end{eqnarray}
where $\Delta_r=\omega_{rg}-\omega_g$ and $\Delta_e=\omega_{eg}-\omega_g+\omega_e$.

We write the state $|\Psi(t)\rangle$ as,
\begin{eqnarray}
|\Psi(t)\rangle=c_g|g\rangle+c_e|e\rangle+c_r|r\rangle.
\end{eqnarray}
From the Schr\"{o}dinger equation, we can obtain 
\begin{eqnarray}
i\dot{c}_g&=&\Omega_gc_r,\\
i\dot{c}_e&=&\Delta_ec_e+\Omega_ec_r,\\
i\dot{c}_r&=&\Delta_rc_r+\Omega_gc_g+\Omega_ec_e.
\end{eqnarray}
We assume that $\Omega_e$ and $\Omega_g$ are real
numbers.

If the detuning $\Delta$ is much greater than the
coupling strengths $\Omega_e,\Omega_g$, then
the upper state can be adiabatically eliminated.
We can obtain 
\begin{eqnarray}
c_r\approx-\frac{\Omega_g}{\Delta_r}c_g-\frac{\Omega_e}{\Delta_r}c_e.
\end{eqnarray}
The effective Hamiltonian can thus be obtained as
\begin{equation}
H_{\rm eff}=-\frac{\Omega^2_g}{\Delta_r}|g\rangle\langle{g}|+\Big(\Delta_e-\frac{\Omega^2_e}{\Delta_r}\Big)|e\rangle\langle{e}|+\Omega(|g\rangle{\langle}e|+{\rm H.c.}),
\end{equation}
where $\Omega=-\Omega_e\Omega_g/{\Delta_r}$ is the effective
coupling strength between the two internal states $|g\rangle$ 
and $|e\rangle$. We assume that $\Omega_g$ and $\Omega_e$ are 
approximately equal and let $\Delta_e=\Delta$ in Eq.~(\ref{Hamint}).


\begin{thebibliography}{99}
\bibitem{Fert}
A. Fert, Rev. Mod. Phys. {\bf 80}, 1517 (2008). 

\bibitem{Nielsen}
M. A. Nielsen and I. L. Chuang, {\it Quantum Computation
and Quantum Information} (Cambridge University Press,
Cambridge, 2000).

\bibitem{Albiez}
M. Albiez, {\it et al.}, Phys. Rev. Lett. {\bf 95},
010402 (2005).

\bibitem{Folling}
S. F\"{o}lling, {\it et al.}, Nature {\bf 448},
1029 (2007).

\bibitem{Bakr}
W. S. Bakr {\it et al.}, Nature {\bf 462} 74 (2009)

\bibitem{Weitenberg}
C. Weitenberg {\it et al.}, Nature {\bf 471}, 319 (2011).

\bibitem{Volz}
J. Volz {\it et al.}, Nature {\bf 475}, 210 (2011).



\bibitem{Ng}
H. T. Ng, C. K. Law, and P. T. Leung, Phys. Rev. A {\bf 68}, 013604 (2003).

\bibitem{Holthaus}
M. Holthaus, Phys. Rev. A, {\bf 64}, 011601(R) (2001);
C. Weiss and T. Jinasundera, Phys. Rev. A {\bf 72}, 053626 (2005);
T. Jinasundera, C. Weiss, M. Holthaus, Chem. Phys. {\bf 322}, 118 (2006).

\bibitem{MMolina}
L. Morales-Molina and J. Gong, Phys. Rev. A {\bf 78}, 041403(R) (2008).

\bibitem{Forster}
L. Forster, {\it et al.}, Phys. Rev. Lett. {\bf 103}, 233001 (2009); 
Q. Beaufils {\it et al.}, {\it ibid.} {\bf 106}, 213002 (2011).

\bibitem{DFrazer}
D. R. Dounas-Frazer, A. M. Hermundstad and L. D. Carr,
Phys. Rev. Lett. {\bf 99} 200402 (2007); P. Cheinet {\it et al.},
{\it ibid.} {\bf 101}, 090404 (2008). 


\bibitem{Creffield}
C. E. Creffield, Phys. Rev. Lett. {\bf 99}, 110501 (2007);
O. Romero-Isart and J. J. Garc\'{i}a-Ripoll, Phys. Rev. A {\bf 76},
052304 (2007); Y. Qian, Ming Gong and C. Zhang, {\it ibid.}
{\bf 84}, 013608 (2011).

\bibitem{Harber}
D. M. Harber {\it et al.}, Phys. Rev. A {\bf 66}, 053616 (2002).

\bibitem{Kuklov}
A. B. Kuklov and B. V. Svistunov, Phys. Rev. Lett. {\bf 90}, 100401 (2003).

\bibitem{Kitagawa}
M. Kitagawa and M. Ueda, Phys. Rev. A {\bf 47}, 5138 (1993).

\bibitem{Wineland} 
D. J. Wineland, J. J. Bollinger, W. M. Itano and D. J. Heinzen, 
Phys. Rev. A {\bf 50}, 67 (1994).


\bibitem{Sorensen}
A. Sorensen, L.-M. Duan, I. Cirac, P. Zoller, Nature {\bf 409}, 63 (2001).


\bibitem{Esteve}
J. Est\`{e}ve {\it et al.}, Nature {\bf 455}, 1216 (2008);
M. F. Riedel {\it et al.}, {\it ibid.} {\bf 464}, 1170 (2010).


\bibitem{Matthews}
M. R. Matthews {\it et al.}, Phys. Rev. Lett. {\bf 81}, 243 (1998).

\bibitem{Milburn}
G. J. Milburn, J. Corney, E. M. Wright and D. F. Walls, Phys. Rev. A
{\bf 55}, 4318 (1997).

\bibitem{Maussang}
K. Maussang {\it et al.}, arXiv:1005.1922..


\bibitem{Steck}
D. A. Steck, {Rubidium 87 D line data}, http://steck.us/alkalidata/ (2010)

\bibitem{Barnett}
S. M. Barnett and P. M. Radmore, {\it Methods in Theoretical 
Quantum Optics} (Oxford University Press, Oxford, 2002).

\bibitem{Beenakker}
C. W. J. Beenakker, Phys. Rev. B {\bf 44}, 1646 (1991).

\bibitem{Messiah}
A. Messiah, {\it Quantum Mechanics} (Dover, New York, 1999).
 
\bibitem{Pepino}
R. A. Pepino, J. Cooper, D. Z. Anderson and M. J. Holland, 
Phys. Rev. Lett. {\bf 103}, 140405 (2009).



\end{thebibliography}
\end{document}